\documentclass[osajnl,twocolumn,showpacs,superscriptaddress,10pt]{revtex4-1} 
\usepackage{amsmath,amssymb,graphicx}

\usepackage{mathtools}
\usepackage{booktabs,multirow}

\begin{document}

\newcommand{\pd}[2]{\frac{\partial #1}{\partial #2}} 
% for partial derivatives
\newcommand{\ket}[1]{\left| #1 \right>} % for Dirac bras
\newcommand{\bra}[1]{\left< #1 \right|} % for Dirac kets
\newcommand{\braket}[2]{\left< #1 \vphantom{#2} \right|
 \left. #2 \vphantom{#1} \right>} % for Dirac brackets
\renewcommand{\b}{\mathbf}
\newcommand{\sech}{\text{sech}}

\title{Multipulse storage and manipulation via solitonic solutions}

\author{Rodrigo Guti\'{e}rrez-Cuevas}
\email{rgutier2@ur.rochester.edu}
\affiliation{Center for Coherence and Quantum Optics, University of 
Rochester, Rochester, New York 14627, USA}
\affiliation{Institute of Optics, University of Rochester, Rochester, New 
York 14627, USA}
\author{Joseph H. Eberly}
\affiliation{Center for Coherence and Quantum Optics, University of 
Rochester, Rochester, New York 14627, USA}
\affiliation{Department of Physics and Astronomy, University of Rochester, 
Rochester, New York 14627, USA \vspace{-.5cm}}

%\date{}%Compiled \today}

%\pacs{42.50.Gy,42.50.Md,42.65.Tg,42.65.Sf}

%

\begin{abstract} \vspace{-.5cm}
Solutions to the Maxwell-Bloch equations for a $\Lambda$ system are 
computed using the single-soliton Darboux transformation and the nonlinear 
superposition principle. These allow complete control of information 
deposited by a signal pulse (with the help of an auxiliary control pulse) in
the coherence of the medium's ground states by injecting sub-sequential 
pulses. Additionally, we study the encoding of two signal pulses and their 
manipulation by a control pulse and show that multipulse storage and 
control are possible as long as the imprints made by encoding the signal 
pulses are sufficiently separated. 
\end{abstract}

%\setboolean{displaycopyright}{true}
%\doi{\url{http://dx.doi.org/10.1364/ao.XX.XXXXXX}}

\maketitle

\section{Introduction\\}

The observation of a solitary canal wave by Scott Russell in 1834 is 
considered the historical origin of soliton studies, but the enormous growth 
of the field began more than a century later. This is due to the discovery 
of new 
methods for solving the nonlinear equations that describe them, such as 
inverse scattering
 \cite{gardner1967method,ablowitz1973nonlinear,lamb1980elements}, 
 the B\"acklund transformation 
\cite{lamb1971analytical,miura1976backlund} and the Darboux transformation 
\cite{gu2006darboux,cieslinski2009algebraic}, to name a few. This phenomenon 
has been increasingly studied in several fields of physics, and 
particularly in optics \cite{kivshar2003optical}. 
McCall and Hahn were the first to observe these solitary optical waves, as 
reported in their famous papers \cite{mccall1967self,mccall1969self} where 
they introduced the concept of self-induced transparency (SIT). Due to the 
coherent interaction of the pulses with the medium, they can propagate 
without attenuation. This is the shape-preserving property 
of solitons that is sometimes used to define them. McCall and Hahn 
also found that the optical pulses tailor their intensity profile so that 
the total pulse area, defined in terms if the Rabi frequency [see Eqs.~(\ref{eq:rabi})] as
\begin{equation}
\theta(x)=\int^{\infty}_{-\infty}\Omega(x,t)dt,
\label{area}
\end{equation}
tends towards the closest even multiple of $\pi$. This is the very 
well-known 
area theorem, which is a consequence of the smoothing effects of Doppler 
broadening \cite{allen2012optical}.
 
Quantum optical systems have always been envisioned as the ideal candidates 
for building reliable quantum memories. This is due to their small 
decoherence and short interaction times \cite{milonni2004fast}. 
Many procedures have achieved light manipulation 
\cite{grobe1994formation} and storage. Light can be slowed down to the point 
where its information is encoded in the medium 
\cite{fleischhauer2000dark,turukhin2001observation} 
and then retrieved as was shown in \cite{liu2001observation}. Some 
other techniques such as a combination of electromagnetic-induced 
transparency (EIT)  and four-wave mixing 
\cite{vudyasetu2008storage,camacho2009four} have been shown to work. The 
downside of these sorts of procedures is that they rely on a slow resonant 
light-atom interaction that is characteristic of EIT 
\cite{harris1997electromagnetically,boller1991observation}. If this 
interaction is 
instead led by short pulses, we can open the door to high-speed control and 
manipulation of light.

\begin{figure}
\centering
\includegraphics[scale=1]{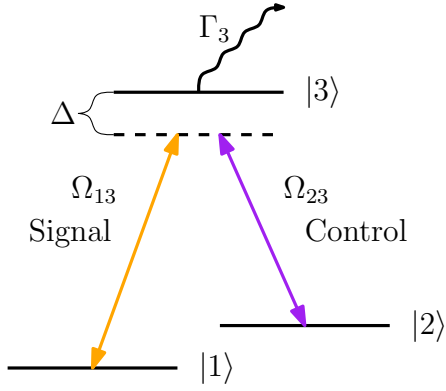}
\caption{\label{lambda} Three level atom in $\Lambda$ configuration, with 
spontaneous emission $\Gamma_3$ from the excited state interacting with two 
fields in two-photon resonance.}
\end{figure}

The regime of broadband pulses interacting with matter leads to new 
possibilities for light control. The interaction of strong electromagnetic 
fields with atomic systems is described by nonlinear evolution equations 
(which are hard to solve) and the use of numerical computation is usually 
required. In some special circumstances, they become integrable and thus 
solvable by the methods previously mentioned. In the particular case of a 
$\Lambda$ system (see Fig.~\ref{lambda}), 
the Maxwell-Bloch equations that define the evolution of the system become 
integrable when both signal and control fields have equal atom-field 
coupling parameters and are in two-photon resonance, as has been shown by 
Park and Shin \cite{park1998matched} and Clader and Eberly 
\cite{clader2007two}. This leads to solitonic solutions even in non-ideal 
media preparation such as ``mixonium'' \cite{clader2008two}, a partially 
coherent medium. Other studies have been carried out for the case of 
ultra-cold atoms, where one can neglect the effects of Doppler broadening 
and homogeneous relaxation. Given these assumptions, Groves \emph{et al.} 
deduced a second order solution that led to an alternative scheme 
for storage and manipulation of light \cite{groves2013jaynes}. Complementary 
numerical simulations \cite{gutierrez2015manipulation} showed the relevance 
of this procedure even under the effects of the ever-present spontaneous 
emission. Here we continue this work by presenting new solutions that allow 
full control of the information stored in the ground state elements of the 
density matrix. We present the generalization to multiple pulse storage as 
well as a three-step control by the corresponding computation of 
higher order solutions.

\section{Theoretical framework\\}

We consider the interaction of strong short pulses with a $\Lambda$ system 
in two-photon resonance with each field tuned to address a different atomic 
transition. The atomic dipole operator is taken to be 
$\boldsymbol{d}=\boldsymbol{d}_{13}\ket{1}\bra{3}+
\boldsymbol{d}_{23}\ket{2}\bra{3}+
\boldsymbol{d}_{31}\ket{3}\bra{1}+\boldsymbol{d}_{32}\ket{3}\bra{2}$, 
thus only linking 
levels 1 to 3 and 2 to 3. As is customary, the fields are written in 
carrier-envelope form:
\begin{align}
\boldsymbol{E}(x,t)=&\boldsymbol{\mathcal E}_{13}
(x,t)e^{i(k_{13}x-\omega_{13}t)} \nonumber \\
&+\boldsymbol{\mathcal E}_{23}(x,t)e^{i(k_{23}x-\omega_{23}t)}
+c.c
\label{carrier}
\end{align}
where $\omega_{13}$ and $\omega_{23}$ are the field frequencies, $k_{13}$ 
and $k_{23}$ are the vacuum wave numbers and 
$\boldsymbol{\mathcal{E}}_{13}(x, t )$ and 
$\boldsymbol{\mathcal E}_{23}(x,t)$ are the 
slowly-varying field envelopes. We 
assume that the pulses are short enough so that we can neglect the effects 
of spontaneous emission but long enough so that the envelopes change slowly 
over many cycles of the optical frequency, thus justifying the slow-varying 
envelope approximation (SVEA).
Following \cite{groves2013jaynes} we refer to the 1-3 field as the signal 
pulse and the 2-3 field as the control pulse. Abandoning the bare 
frequencies in favor of the common detuning, the total Hamiltonian in the 
rotating wave approximation (RWA) takes the form:
\begin{equation}
 H=-\frac{\hbar}{2} \left(
\begin{array}{ccc} 
0&0&\Omega_{13}^*\\
0&0&\Omega_{23}^*\\
\Omega_{13}&\Omega_{23}&-2\Delta
\end{array}
\right)
\label{hrwa}
\end{equation}
where we defined the Rabi frequencies,
\begin{subequations}
\label{eq:rabi}
\begin{align}
\Omega_{13}
(x,t)&=2\boldsymbol{d}_{31}\cdot 
\boldsymbol{\mathcal E}_{13}(x, t )/\hbar,\\
\shortintertext{and}
\Omega_{23}
(x,t)&=2\boldsymbol{d}_{32}\cdot 
\boldsymbol{\mathcal E}_{23}(x, t )/\hbar,
\end{align}
\end{subequations}
and the detuning $\Delta=(E_3-
E_1)/\hbar-\omega_{13}=(E_3-E_2)/\hbar-\omega_{23}$ (here $E_i$ denotes the 
energy of level $\ket i$).
The atomic system evolves according to the von Neumann equation for the 
density matrix:
\begin{equation}
i\hbar \pd{\rho}{t}=[H, \rho]
\label{neumann}
\end{equation}
and the fields follow Maxwell's wave equation in the SVEA:
\begin{subequations}
\label{meqs}
\begin{align}
\left(\pd{ }{x}+\frac{1}{c}\pd{}{t}\right)\Omega_{13}&=i\mu_{13} \rho_{31}\\
\shortintertext{and}
\left(\pd{ }{x}+\frac{1}{c}\pd{}{t}\right)\Omega_{23}&=i\mu_{23} \rho_{32}.
\end{align}
\end{subequations}
Here, we defined the atom-field coupling parameters 
$\mu_{j3}=N\omega_{j3}|d_{j3}|^2/\hbar \epsilon_0c$ with $j=1,\,2$.   
This gives us a set of eight nonlinear partial differential equations that 
need to be solved simultaneously. 
As stated previously, we need to consider the special case of two-photon 
resonance and equal atom-field coupling parameters, $\mu_{13}=\mu_{23}=\mu$. 
This way we can use the methods described in the introduction. In the 
travelling-wave coordinates $T=t-x/c$ and $Z=x$, Eqs.~(\ref{neumann}) and 
(\ref{meqs}) take the form: 
\begin{subequations}
\begin{equation}
i\hbar \pd{\rho}{T}=[ H, \rho] \label{neumann2}
\end{equation}
and
\begin{equation}
\pd{H}{Z}=-\frac{\hbar \mu}{2}[W, \rho],
\end{equation}
\label{mb}
\end{subequations}
where the constant matrix
\begin{equation}
 W=i\ket{3}\bra 3= \left(
\begin{array}{ccc}
0&0&0\\
0&0&0\\
0&0&i
\end{array}
\right)
\end{equation}
has been introduced.
By combining these two equations it is easily shown that the Lax equation, 
\begin{equation}
\partial_Z  U-\partial_T V +[U,V]=0,
\end{equation}
is satisfied where the Lax operators are defined as $U=-(i/\hbar) H - 
\lambda W$ and $ V=(i\mu/2\lambda) \rho$, and $\lambda$ is a constant known 
as the spectral parameter. This effectively shows that the Maxwell-Bloch 
equations [Eqs.~(\ref{mb})] are integrable.

Throughout this framework, we have only considered the longitudinal spatial 
dimension; this was justified by the assumption of plane waves. In reality, 
the pulses will have a nonuniform intensity profile that will inevitably 
lead to deviation from the theoretical assumption by effects such as pulse 
stripping, diffraction and self-focusing \cite{mccall1969self}. To 
mitigate this we can use pulses that have a coherent frequency profile, 
have their bandwidth determined only by the finite extent of their 
envelope, and are free of chirping. Additionally, an aperture can be used to 
ensure a planar wavefront and homogeneous distribution of intensity. An 
example of this can be seen in the beautiful experiments on SIT by Gibbs and 
Slusher \cite{gibbs1970peak,slusher1972self}. Another thing worth 
mentioning is that the finite extent of the pulses imposes some 
restrictions in the use of atomic beams to get rid of Doppler broadening, 
namely, the transit time broadening must be smaller than the bandwidth of 
the pulses. This must be satisfied in order to  ensure that the atoms where 
the signal pulse was encoded are the same that are interacting with the 
subsequent pulses.\\

\section{Solution method\\}

\subsection{Darboux transformation\\}

\begin{figure*}[t]
\centering
\includegraphics[scale=1]{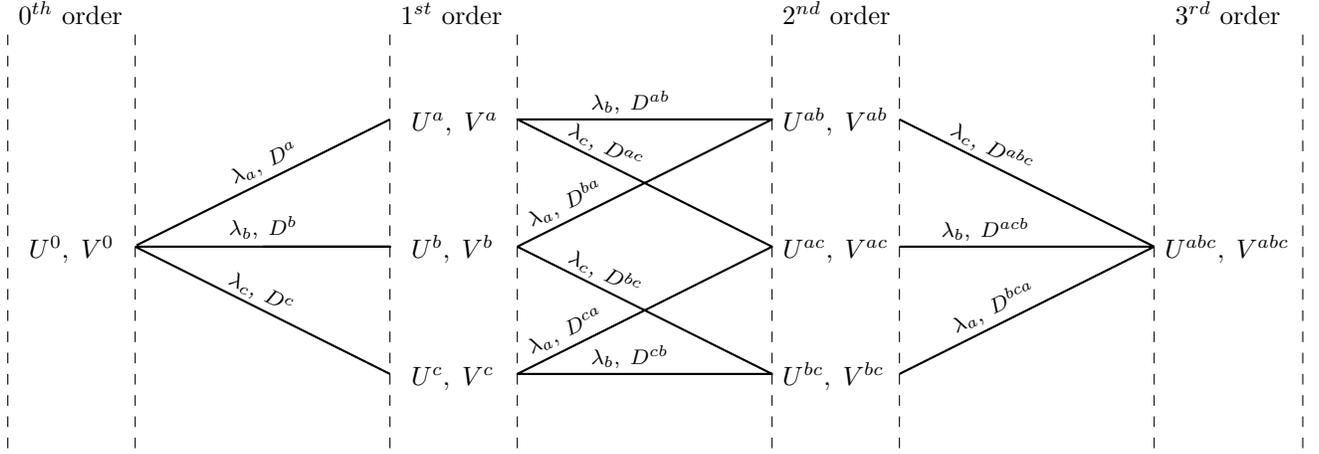}
\caption{\label{bianchi} Bianchi diagram for the theorem of permutability.
By requiring commutativity of the diagram we are able to find an algebraic 
method to compute higher order solutions.}
\end{figure*}
The basic idea of the Darboux transformation is to start from a system of 
partial differential equations of the form:
\begin{equation}
\label{psi0}
\pd{\psi}{T}=U \psi \quad \text{and} \quad \pd{\psi}{Z}= V \psi ,
\end{equation}
then consider a transformation $\bar{\psi}= D \psi$ so that
\begin{equation}
\pd{\bar \psi}{T}=\bar U \bar \psi \quad \text{and} \quad \pd{\bar \psi}
{Z}=\bar V \bar \psi,
\end{equation}
where 
\begin{subequations}
\begin{align}
\bar U=DUD^{-1}+(\partial_T D)D^{-1} \\
\shortintertext{and}
\bar V=DVD^{-1}+(\partial_Z D)D^{-1}.
\end{align}
\label{puv}
\end{subequations}
Now we want to construct the operator $D$ in terms of known parameters such 
as the original solution to the linear Eq.~(\ref{psi0}). We also want 
it to preserve the spectral dependence of the Lax pair, in which case we 
call it a Darboux matrix. Additionally, we need to conserve the hermiticity
of the density matrix and Hamiltonian which in turn requires that 
$U(\lambda)^\dag=-U(\lambda^*)$ and $V(\lambda)^\dag=-V(\lambda^*)$. 
Such Lax 
operators belong to what is known as the unitary reduction. 
Starting from the ``single-soliton'' Darboux matrix, we have that
\begin{equation}
D=(\lambda-\lambda_1)I-(\lambda_1-\lambda_1^*)P
\end{equation}
in order to satisfy all the required properties 
\citep{cieslinski2009algebraic}, where $I$ is the identity matrix and $P$ is 
a hermitian projection operator
($P^2=P$ and $P=P^\dag$). Taking $\lambda_1$ to be complex leads to more 
complicated solutions and so we will consider it to be purely imaginary, 
thus getting
\begin{equation}
D=\lambda I +\lambda_1 (2P-I).
\label{D}
\end{equation}

From the previous section we know that the Lax operators for the 
Maxwell-Bloch equations [Eqs.~(\ref{mb})] have the following spectral 
dependence:
\begin{subequations}
\begin{align}
U(\lambda)&=\lambda^0 U_0+\lambda^1 U_1\\
V(\lambda)&=\lambda^{-1} V_{-1}.
\end{align}
\end{subequations}
After inserting this into Eqs.~(\ref{puv}) and collecting terms of equal 
order in 
$\lambda$ we find:
\begin{subequations}
\begin{align}
\lambda^0:&\quad  \bar U_0 M=M U_0+\partial_T M \\
\lambda^1:&\quad \bar U_0 +\lambda_1 \bar U_1 M=U_0 +\lambda_1 M U_1 \\
\lambda^2:&\quad \bar U_1 =U_1 \label{eqW}
\end{align}
\label{um}
\end{subequations}
and
\begin{subequations}
\label{vm}
\begin{align}
\lambda^{-1}:&\quad  \bar V_{-1} M=M V_{-1} \\
\lambda^0:&\quad \bar V_{-1} =V_{-1} +\lambda_1\partial_Z M
\end{align}
\end{subequations}
where we defined the unitary involution $M=2P-I$. It is worth noting that 
from Eq.~\eqref{eqW} it is clear that the matrix $W$ will remain constant between 
solutions, as it should. Writing the projection operator as 
$P=\ket \varphi \bra \varphi / \braket{\varphi}{\varphi}$ it is possible to 
use Eqs.~(\ref{um}) and (\ref{vm}) to determine a set of equations for the 
column vector $\ket \varphi$. By means of the properties of the projection 
operator, the result can be simplified to obtain a set of two linear 
differential equations that determine $\ket \varphi$: 
\begin{subequations}
\begin{align}
\left(I \partial_T-U(-\lambda_1)\right)\ket \varphi =0, \\
\left(I \partial_Z-V(-\lambda_1)\right)\ket \varphi =0.
\end{align}
\label{eqphi}
\end{subequations}
This derivation is similar to the one presented by Clader and Eberly in 
\cite{clader2007two}. Solving Eqs.~(\ref{eqphi}) determines the projection 
operator and thus the Darboux matrix. 

From the definition of the Lax operators for the Maxwell-Bloch equations we 
can relate the new solution to the first order density matrix and 
Hamiltonian. If we assume that our seed solution was given by the density 
matrix $\rho^0$ and the Hamiltonian $H^0$, it follows that
\begin{subequations}
\begin{equation}
H=H^{0}-i\hbar \lambda_1\left[ M,W \right],
\end{equation}
\begin{equation}
\rho=M \rho^0 M.\label{mden}
\end{equation}
\end{subequations}

\subsection{Nonlinear superposition rule\\}

In principle, the method described in the previous section could be used to 
compute higher order solutions, but the reality is that Eqs.~(\ref{eqphi}) 
become harder to solve with each step. Luckily there is a much simpler
way to achieve this: the theorem of permutability (Fig.~\ref{bianchi}). 
Starting from a seed solution (zeroth order) 
$U^0$ and $V^0$, it is possible to construct two new solutions $U^a$ and 
$V^a$ with the associated parameter $\lambda_a$ and $U^b$, and $V^b$ with 
associated parameter $\lambda_b$ using the Darboux matrices $D^a$ and $D^b$ 
that are of the form given by Eq.~\eqref{D}. From these first order 
solutions we can construct second order solutions by applying $D^{ab}$ with 
parameter $\lambda_b$ to the $a$ solution and  $D^{ba}$ with parameter 
$\lambda_a$ to the $b$ solution thus obtaining the new pairs of Lax 
operators $U^{ab}$ $V^{ab}$ and $V^{ba}$ $U^{ba}$. The theorem of 
permutability asserts that there is nothing special about the order in which 
the second order solutions are computed. This is equivalent to requiring 
commutativity of the Bianchi diagram as shown in Fig.~\ref{bianchi}. Both 
second order solutions should then be the same, $U^{ab}=U^{ba}$ and 
$V^{ab}=V^{ba}$.

Using Eqs.~(\ref{um}) and (\ref{vm}) for the second order Lax pair and 
setting the result from the two paths equal to each other, we can derive the 
following expression for the second order involution matrix:
\begin{equation}
M^{ab}=(\lambda_a M^a-\lambda_b M^b)(\lambda_a M^a M^b -\lambda_b I)^{-1}.
\label{nls}
\end{equation}
It is easy to relate this to the density matrix and Hamiltonian, thus 
bypassing the need to compute the Darboux matrix and solve complicated 
differential equations. Using the properties of the involution matrices it 
is easy to show that
\begin{subequations}
\begin{equation}
\label{denab}
\rho^{ab}=M^{ab}M^a\rho^0 M^a M^{ab}
\end{equation}
and
\begin{equation}
H^{ab}=H^0-i \hbar 
(\lambda_a^2-\lambda_b^2)\left[(\lambda_aM^a-\lambda_bM^b),W\right].
\end{equation}
\end{subequations}

This treatment can be extended to obtain third order solutions by purely 
algebraic methods. Once more we assume the 
commutativity of the Bianchi diagram 
up to the third order which, again, is a statement of the independence of 
the path taken from a total of six possibilities (see Fig.~\ref{bianchi}). 
After some simplifications using the properties of the involution matrices 
and the solution for the second order, the expression for the third order 
involution matrix can be written in similar form as the one for the second 
order, namely:
\begin{equation}
\label{nls3}
M^{abc}=(\lambda_b M^{ab}- \lambda_c M^{ac})(\lambda_b 
M^{ab}M^{ac}-\lambda_c I)^{-1}.
\end{equation}
This is the same as using the nonlinear superposition rule [Eq.~\eqref{nls}] on 
two second order solutions. Finally, relating this result to the density 
matrix and Hamiltonian we have
\begin{subequations}
\begin{equation}
\rho^{abc}=M^{abc}M^{ab}M^a\rho^0 M^a M^{ab}M^{abc}
\end{equation}
and
\begin{equation}
H^{abc}=H^a-i \hbar 
(\lambda_b^2-\lambda_c^2)\left[(\lambda_bM^{ab}-\lambda_cM^{ac}),W\right].
\end{equation}
\end{subequations}

{
\renewcommand{\arraystretch}{1.7}
\setlength{\tabcolsep}{5pt}
\begin{table*}
\centering
\caption{\label{Ma}  Elements of the involution matrix $M^a$ for the three 
types of first order solutions considered.
\footnote{
For type 1 we only show the 
elements in the limits of infinite negative and positive times as these can 
be written in a simple form. We define the parameters 
$A_{jk}=a_ja_k^*/|a_ja_k|$ and $\eta_{jk}^a=\ln |a_j/a_k|$ that control the 
phase and location of the pulses.}
} 
\begin{tabular}{ccccc}
\hline
\hline
& \multicolumn{2}{c}{Type 1 ($a_1,a_2,a_3\neq 0$)}& Type 2 ($a_1=0$)& Type 3 
($a_2=0$)\\
 & $T/\tau_a \ll-1$ & $T/\tau_a \gg1$ & for all times & for all times\\
 \hline
$M_{11}^a$& $\tanh \left(\frac{T}{\tau_a} -\frac{\mu \tau_a}{2} 
Z+\eta_{13}^a\right)$ & $\tanh \left( -\frac{\mu \tau_a}{2} 
Z+\eta_{12}^a\right)$ &$ -1$ &  $\tanh \left(\frac{T}{\tau_a} -\frac{\mu 
\tau_a}{2} Z+\eta_{13}^a\right)$\\
$M_{22}^a$& $-1$ &$-\tanh \left( -\frac{\mu \tau_a}{2} 
Z+\eta_{12}^a\right)$& $\tanh \left(\frac{T}{\tau_a}+\eta_{23}^a\right)$& 
$-1$\\
$M_{33}^a$&$-\tanh \left(\frac{T}{\tau_a} -\frac{\mu \tau_a}{2} 
Z+\eta_{13}^a\right)$ & $-1$ & $-\tanh \left(\frac{T}
{\tau_a}+\eta_{23}^a\right)$ & $-\tanh \left(\frac{T}{\tau_a} -\frac{\mu 
\tau_a}{2} Z+\eta_{13}^a\right)$\\
$M_{12}^a$&$0$&$ A_{12}\sech\left( -\frac{\mu \tau_a}{2} Z+\eta_{12}^a 
\right) $ & $0$&$0$\\
$M_{13}^a$& $ A_{13}\sech\left( \frac{T}{\tau_a}-\frac{\mu \tau_a}{2} 
Z+\eta_{13}^a \right) $ & $0$ & $0$ &  $ A_{13}\sech\left( \frac{T}
{\tau_a}-\frac{\mu \tau_a}{2} Z+\eta_{13}^a \right) $ \\
$M_{23}^a$ &$0$& $ A_{23}\sech\left(\frac{T}{\tau_a} +\eta_{23}^a \right) $ 
&$ A_{23}\sech\left(\frac{T}{\tau_a} +\eta_{23}^a \right) $ & $0$\\
\hline
\hline
\end{tabular}
\end{table*}
}

\section{First order solutions\\}

After reviewing the solution method, we now proceed to solve the 
Maxwell-Bloch equations [Eqs.~(\ref{mb})]. For simplicity we 
will consider the 
special case of zero detuning. The seed is taken to be the trivial solution 
of a quiescent medium ($\rho^0=\ket 1 \bra 1 $) and no fields 
($\Omega_{13}=\Omega_{23}=0$), so that $H^0=0$. It is easy to see 
that by solving Eqs.~(\ref{eqphi})
\begin{equation}
\ket{\varphi^a}=
\left(\begin{array}{c}
a_1 e^{-\mu \tau_a Z/2}\\
a_2\\
a_3 e^{-T/\tau_a}
\end{array}
\right),
\end{equation}
where $a_1$, $a_2$, $a_3$ are constants of integration. 
Here, we wrote the Darboux parameter as $\lambda_a=i/\tau_a$, with 
$\tau_a \in \mathbb{R}$ which makes it easier to associate it to a physical
quantity, namely, the duration of the pulses.
Two possible solutions arising from this have already been thoroughly 
studied by Groves, Clader and Eberly \citep{groves2013jaynes} so we will 
just summarize the results that we will need. Table \ref{Ma} contains all 
the elements of the involution $M^a$ for the three first order 
solutions that are going to be considered.

\begin{figure*}
\centering
\includegraphics[scale=1]{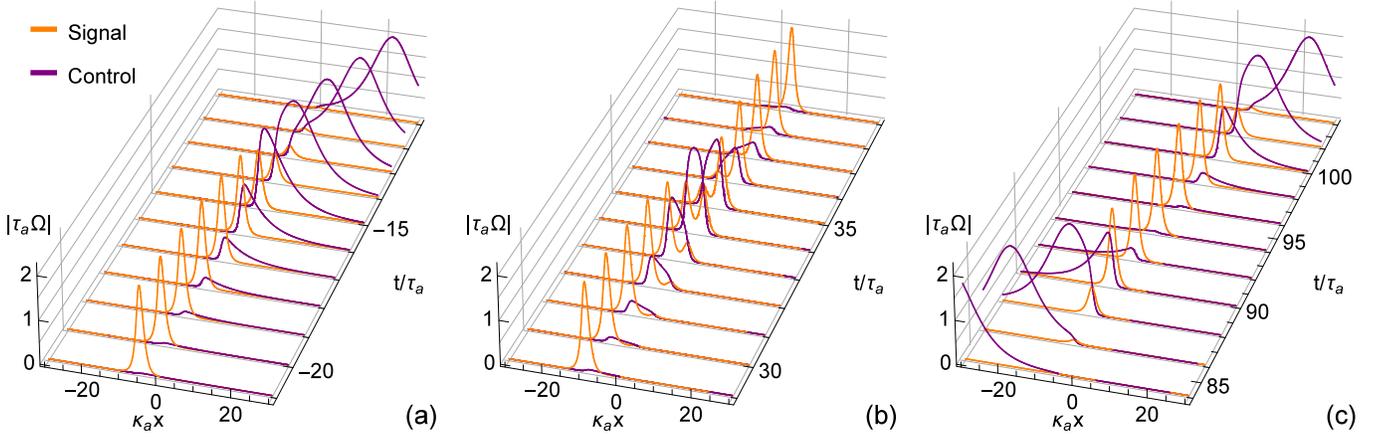}
\caption{\label{pulse1} Third order solution for a single 
imprint obtained from the nonlinear superposition of a first order solution 
of each type: (a) shows the initial encoding of the signal pulse at 
$\kappa_ax_1=0$, (b) shows the collision of a second signal pulse with the 
imprint which displaces it to $\kappa_ax_1=-5$ and (c) shows the collision 
of the imprint with a control pulse which moves it to $\kappa_ax_1=5$. The 
corresponding imprints are depicted in Fig.~\ref{den1}.}
\end{figure*}

The most general solution (type 1) is given by taking all the integration 
constants to be different from zero. In this case we end up with a special 
case of the two-pulse soliton solution previously 
found by Clader and Eberly 
in \citep{clader2007two}. In the limit $t/\tau_a\ll-1$ we have a SIT-like 
signal pulse propagating, driving population from the ground state $\ket 1$ 
into the excited state $\ket 3$ and coherently driving it back, thus 
obtaining the characteristic SIT $2\pi$-pulse shaped as an hyperbolic 
secant. As  the control pulse is only zero in the limit of infinite 
negative time, some of the excited population is coherently driven into the 
ground state $\ket 2$, thus amplifying the seed of the control pulse. Its 
effect slowly takes over, up to the point where the signal pulse starts to 
be depleted as the control pulse is amplified [see Fig.~\ref{pulse1}(a)].

\begin{figure}[t]
\centering
\includegraphics[scale=1]{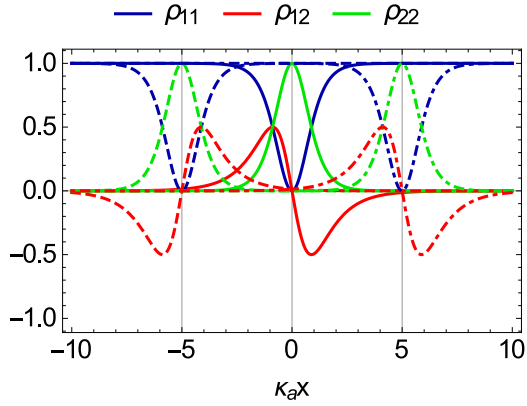}
\caption{\label{den1} Information encoding and control in a 
$\Lambda$ system:
Imprint as it has been encoded in the ground state density
matrix elements after the initial encoding (continuous line), after the 
first backward 
displacement (dashed lines) and after the second forward displacement 
(dot-dashed lines). The imprint was generated and displaced by the
pulse sequence depicted in Fig.~\ref{pulse1} and the snapshots were taken
at times $t/\tau_a = 0$, $60$ and $150$, respectively.}
\end{figure}

During this transfer the signal pulse encodes its information into the 
ground state elements $\rho_{11}$, $\rho_{22}$ and $\rho_{12}$ of the 
density matrix. Table \ref{Ma} only shows the elements of the matrix $M^a$ 
at infinite positive and negative times as these can be written in a simple 
form and are the most relevant information. From these and Eq.~\eqref{mden} 
the shape of the ground state elements of the density matrix can be obtained 
and are given by:
\begin{subequations}
\label{dena}
\begin{equation}
\rho_{11}^a=\tanh^2 \left( -\frac{\mu \tau_a}{2} Z+\eta_{12}^a\right),
\end{equation}
\begin{equation}
\rho_{12}^a= A_{12} \sech \left( -\frac{\mu \tau_a}{2} 
Z+\eta_{12}^a\right)\tanh \left( -\frac{\mu \tau_a}{2} Z+\eta_{12}^a\right),
\end{equation}
\begin{equation}
\rho_{22}^a=\sech^2 \left( -\frac{\mu \tau_a}{2} Z+\eta_{12}^a\right).
\end{equation}
\end{subequations}
All other elements of $\rho^a$ are zero. The location of the imprint is 
where the population of state $\ket 2$ has a maximum (this also corresponds 
with the minimum of $\rho_{11}^a$ and the zero of $\rho_{12}^a$) and thus is 
given by 
\begin{equation}
\label{ximp}
\kappa_a x_1^a=\eta^a_{12},
\end{equation}
where $\kappa_a=\mu \tau_a /2$ is the absorption coefficient in the absence 
of Doppler broadening. An example of this imprint is depicted by the plots 
with continuous lines in Fig.~\ref{den1}.
The addition of Doppler broadening would affect the definition of the 
absorption coefficient and thus change the group velocity of the pulses in 
the medium, but the encoding would still carry through. It is also worth 
noting that while the two pulses are active, the 
area of the individual pulses is no longer equal to $2\pi$ but the total 
pulse area as defined in \citep{clader2007two},
\begin{equation}
\theta_\text{tot}=\sqrt{|\theta_{13}|^2+|\theta_{23}|^2},
\label{atot}
\end{equation}
remains constant and equal to $2\pi$. After the storage process is over, 
$t/\tau_a\gg1$, we have a $2\pi$-control pulse propagating away at the 
speed of light as it is decoupled from the medium. Both signal and control 
pulses have a duration of $\tau_a$ and are time-matched.

Another possibility is taking one of the integration constants to be zero. 
If $a_1=0$ then we obtain a type 2 solution which is a $2\pi$-control pulse 
traveling at the speed of light completely decoupled from the medium. If 
instead we take $a_2=0$ we end up with an SIT solution for the signal pulse 
which we will refer to as type 3. This pulse propagates at a reduced group 
velocity, coherently driving population from state $\ket 1$ to the excited 
state $\ket 3$ and back again, thus keeping its hyperbolic secant shape as 
it propagates.\\

\section{Single imprint manipulation\\}
\label{singimp}

\subsection{Backward-transfer solution\\}
\label{backtransf} 

Now we make use of the nonlinear superposition rule [Eq.~\eqref{nls}] 
to combine a 
type 1 solution with a type 3 solution, to which we assign the letters $a$ 
and $b$ respectively. The parameters $\eta_{jk}$ defined in Table \ref{Ma} 
did not have any relevance other than to control where the signal pulse 
deposited its information into the medium for the type 1 solutions. Now 
that we are superimposing two first order solutions it acquires renewed 
relevance as it  also controls the order of the pulses and whether the type 
1 pulse has enough time to make the imprint before the type 3 pulse collides 
with it. 
The Rabi frequencies in the limit of infinitely negative time are given by
\begin{subequations}
\label{rabiab}
\begin{align}
\Omega_{13}^{ab}=&-\frac{2i}{\tau_a}A_{13}^*\sech \left( \frac{T}{\tau_a} 
-\frac{\mu \tau_a}{2} Z+\eta_{13}^a+ \delta^{ab} \right) \nonumber \\
&+\frac{2i}{\tau_b}B_{13}^*\sech \left( \frac{T}{\tau_b} -\frac{\mu \tau_b}
{2} Z+\eta_{13}^b-	 \delta^{ab} \right),
\end{align}
\begin{equation}
\Omega_{23}^{ab}=0,
\end{equation}
\end{subequations}
where we defined the phase lag parameter
\begin{equation}
\delta^{ab}=\ln \left| \frac{\tau_a+\tau_b}{\tau_a-\tau_b}\right|.
\end{equation}
Here, we will stick to the case where $\eta_{13}^a\gg\eta_{13}^b$ to 
guarantee that the signal pulse from the type 1 solution has enough time to 
encode its information into the medium before the second signal pulse 
collides with it. 

From the type 1 solution, we have that the first signal pulse will imprint 
its information at a location determined by Eq.~\eqref{ximp}. This proceeds 
as the already described type 1 first order solution, and the corresponding 
pulse dynamics are shown in Fig.~\ref{pulse1}(a). Then comes the second signal 
pulse of different duration $\tau_b$. As it approaches the imprint, the 
second signal pulse starts to decay as it gives way to a control pulse, 
which mediates the transfer of the peak of the signal pulse from the new 
location of the imprint to where it was first made. Therefore the imprint is 
effectively pushed backwards. Finally, the control pulse decays and the 
signal pulse continues to propagate as an SIT-type solution. This process is 
shown in Fig.~\ref{pulse1}(b). This effect is clearly due to the long tails of 
the sech-shaped pulses that can sense changes in the medium long before the 
peak of the pulses and thus interact with it accordingly. This is similar to 
what happens with fast light, where the long tails of the pulse sense the 
inverted 
medium and so  the peak of the pulse is displaced 
at a speed greater than that of light due to stimulated emission
\cite{wang2000gain,boyd2002slow,clader2006coherent,clader2007theoretical}. 

In order to determine what the effect was on the imprint, we need to 
compute the ground state elements of the density matrix in the limit of 
infinitely positive long time using Eq.~\eqref{denab}. We find that
\begin{subequations}
\begin{equation}
\rho_{11}^{ab}=\tanh^2 \left( -\frac{\mu \tau_a}{2} 
Z+\eta_{12}^a-\delta^{ab}\right),
\end{equation}
\begin{align}
\rho_{12}^{ab}=&-\phi A_{12} \sech \left( -\frac{\mu \tau_a}{2} 
Z+\eta_{12}^a-\delta^{ab}\right)\nonumber\\
&\times\tanh \left( -\frac{\mu \tau_a}{2} Z+\eta_{12}^a-\delta^{ab}\right),
\end{align}
\begin{equation}
\rho_{22}^{ab}=\sech^2 \left( -\frac{\mu \tau_a}{2} 
Z+\eta_{12}^a-\delta^{ab}\right),
\end{equation}
\end{subequations}
where $\phi=\text{sgn}(\tau_a-\tau_b)$ which determines the phase of the 
coherence. 

The effect is clear: The imprint is displaced to the left by an amount 
determined by the phase lag parameter with a possible phase shift in the 
coherence $\rho_{12}$ depending on the relation between the duration of the 
pulses. The dashed line plots in Fig.~\ref{den1} show the displaced imprint. 
This is the same result as the one obtained when superimposing a type 1 and 
a type 2 first order solution except that the sign of the displacement is 
inverted, and so the imprint is moved to the right. This has already been 
shown in \cite{groves2013jaynes,gutierrez2015manipulation}.

\subsection{Multi-step manipulation\\}

Having identified all the relevant parameters in the control of a single 
imprint by studying the second order solutions, we can proceed to extend 
this control to multi-step processes. Here we consider only third 
order solutions which are composed of three steps. The first will be the 
imprinting step, then we can consider a combination of other control and/or 
signal pulses to move the imprint back and forth.

For clarity lets consider the case of pushing the imprint to the left and 
then to the right by means of type 3 and type 2 first order solutions, 
respectively. We compute the third order solution by means of the 
superposition rule given by Eq.~\eqref{nls3}. This situation is depicted in 
Fig.~\ref{pulse1} where each frame corresponds to a step and each will be 
labelled by their corresponding letters $a$, $b$ and $c$. First, the 
information of the initial signal pulse of duration $\tau_a$ is deposited in 
the form of an imprint that is made into the medium as determined by 
Eq.~\eqref{ximp}. Then comes a second signal pulse of duration $\tau_b$. Its 
effect is the same as described in the previous section. The imprint is 
moved to the left and its new location is $\kappa_a 
x_2^a=\kappa_ax_1^a-\delta^{ab}$. As has already been mentioned, there is a 
phase shift in the coherence $\rho_{12}$ if $\tau_b<\tau_a$. Finally, for 
the third step, a control pulse of duration $\tau_c$ comes in. Upon 
interaction with the imprint, it reads the information stored and retrieves 
the initial signal pulse which, in turn, is restored in a new location given 
by $\kappa_a x_3^a=\kappa_a x_2^a+\delta^{ac}$ with the same possibility 
for another $\pi$-phase shift for $\rho_{12}$.
The results of creation and displacements of the imprint
are shown in Fig.~\ref{den1}. Continuous lines show the first imprinted 
density matrix elements, dashed lines represent the first 
displacement to the left and dot-dashed lines show the imprint when it 
is displaced to the right.

It should be clear that the generalization for single imprint manipulation 
to an n$^\text{th}$ order solution follows from the previous results. We can 
readily write the final location of the imprint,
\begin{equation}
\kappa_a x_n^a= \eta_{12}^a-\sum_{i=1}^{m_2} 
\delta^{ab_{i}}+\sum_{i=1}^{m_3} \delta^{ac_i},
\end{equation}
where $m_2$ and $m_3$ are, respectively, the number of type 2 and 3 first 
order solutions that compose this n$^\text{th}$ order solution. 
Additionally, if there is an odd number of pulses with duration shorter than 
$\tau_a$ (the original signal pulse) then there is a $\pi$-phase shift for 
the imprint. \\

\section{Multiple imprint control\\}

\subsection{Two imprints solution\\}

The first step to generalize this control to multiple imprint dynamics is to 
study the second order solution born out of the superposition of two type 1 
solutions. This will simulate the scenario of having two signal pulses each 
with their own control pulse seed. As each pair of pulses have different 
time duration, one of them will be traveling faster. The parameters have 
to be carefully chosen so that the first signal pulse, which we will label 
with the letter $a$, deposits its information into the medium before the 
second, labeled with the letter $b$, catches up. In the limit of infinite 
negative time the Rabi frequencies are given the same expressions as for the 
backward-transfer solution Eqs.~(\ref{rabiab}).  Therefore, in order to have 
the correct order for the pulses, the condition $\eta_{13}^a\gg\eta_{13}^b$ 
must be satisfied.

Here again, after the first pair of pulses of time duration $\tau_a$ has 
deposited the information of the signal pulse, the density matrix elements 
are given by Eqs.~(\ref{dena}) and the location of the first imprint $x_1^a$ 
is determined by Eq.~\eqref{ximp}. When the second imprint is made, if we 
want to preserve the information of each pulse separately, we need to have 
that $|\eta_{12}^a-\eta_{12}^b|\gg1$. Using this and the expressions given 
in Table \ref{Ma} in the infinite time limit, we can work out expressions 
for the density matrix elements around each imprint. 
In the vicinity of the 
first imprint (the one made by the pulses of duration $\tau_a$) we have:
\begin{figure*}
\centering
\includegraphics[width=.9\linewidth]{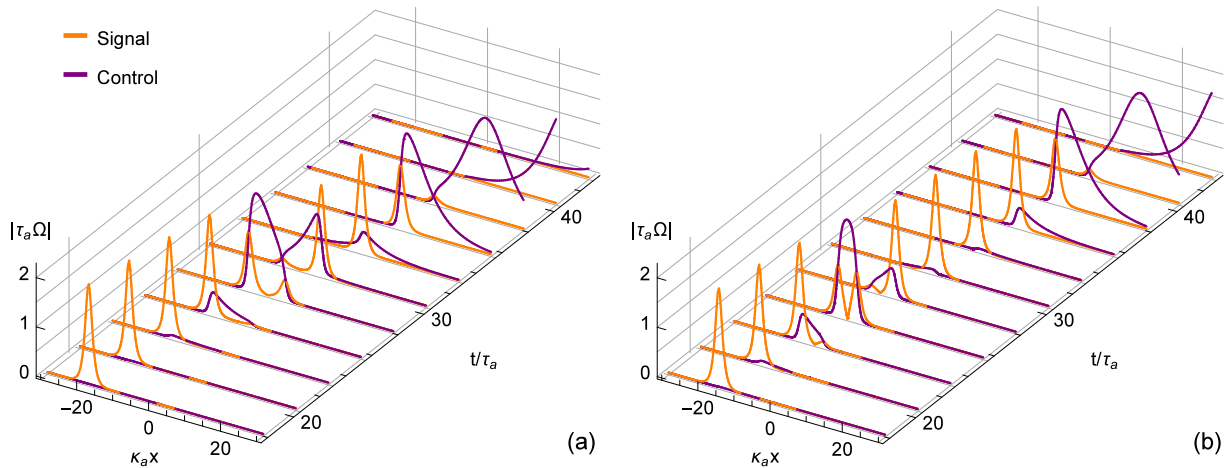}
\caption{\label{pulse2} Encoding of a second signal pulse. 
(a) A signal pulse of duration $\tau_a$ was encoded at $\kappa_a x_1^a=
\eta_{12}^a$ and then comes a second signal pulse of duration $\tau_b$ meant 
to be stored at $\kappa x_1^b=\eta_{12}^b$, with $x_1^a>x_1^b$. (b) Here the 
reverse process is presented, the pulse $\tau_b$ is the first to be encoded
and the pulse $\tau_a$ is the second. The values for the $\eta_{12}$'s are 
kept the same as well as those of the $\tau$'s. Both cases lead to the same 
imprint which is shown in Fig.~\ref{den2}.}
\end{figure*}
\begin{subequations}
\begin{equation}
\rho_{11}^{ab}=\tanh^2 \left( -\frac{\mu \tau_a}{2} Z+\eta_{12}^a+\sigma 
\delta^{ab}\right),
\end{equation}
\begin{align}
\rho_{12}^{ab}=&-\phi A_{12} \sech \left( -\frac{\mu \tau_a}{2} 
Z+\eta_{12}^a+\sigma \delta^{ab}\right)\nonumber\\
&\times\tanh \left( -\frac{\mu \tau_a}{2} Z+\eta_{12}^a+\sigma 
\delta^{ab}\right),
\end{align}
\begin{equation}
\rho_{22}^{ab}=\sech^2 \left( -\frac{\mu \tau_a}{2} Z+\eta_{12}^a+\sigma 
\delta^{ab}\right),
\end{equation}
\end{subequations}
where we defined $\sigma=\text{sgn} (\eta_{12}^a-\eta_{12}^b)$. Around the 
second imprint we have similar expressions with the signs in front of 
$\sigma$ and $\phi$ reversed.

\begin{figure}[t]
\centering
\includegraphics[scale=1]{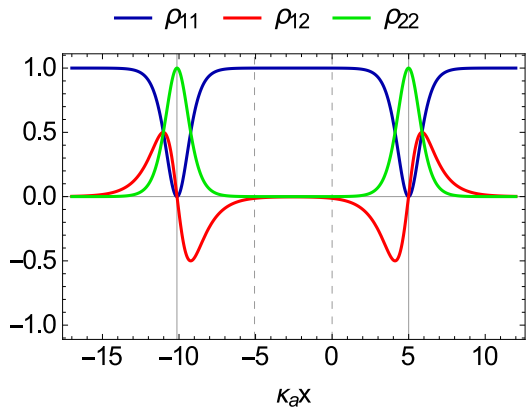}
\caption{\label{den2} Multi-pulse storage in a 
$\Lambda$-system: Each imprint made by the storage of two signal pulses 
of duration $\tau_a$ and $\tau_b$. The 
vertical continuous lines show the location of 
each imprint and the dashed ones show where they would have been encoded if 
they had been by themselves.}
\end{figure}

There are several things to comment about these expressions but the most 
obvious and somewhat shocking is that the parameters $\eta_{13}^a$ and 
$\eta_{13}^b$ are nowhere to be found. This means that the temporal order in 
which the imprints were made is irrelevant; the only thing that matters is 
the spatial ordering as it is shown by the dependence on $\eta_{12}^a$ and 
$\eta_{12}^b$. Therefore the two processes depicted in Fig.~\ref{pulse2} lead 
to the same result. On one hand, we have the situation assumed here, namely, 
that the signal pulse $\tau_a$ stores its information first and 
$x_1^a>x_1^b$. When the second signal pulse $\tau_b$ is coming in, it 
senses the presence of the first imprint and thus ``knows'' that it 
must deposit its information before the value predicted by 
Eq.~\eqref{ximp}. This again is clearly a feature of the long 
tails of the sech-shaped pulses that start interacting 
with the imprint long before the peak collides with it. The 
information is then encoded in a process similar to the one described by a 
type 1 first order solution. The control pulse that comes out from the 
encoding process then pushes the first imprint in a way similar to the 
prediction from a superposition of type 1 and 2 first order solutions. This 
process is shown in Fig.~\ref{pulse2}(a). Now, on the other hand, we have the 
situation where the signal pulse $\tau_b$  encodes its information into the 
medium first. Then comes the second signal pulse $\tau_a$. When it comes 
close to the imprint it acts as a backward-transfer solution, thus displacing 
the imprint by the amount discussed in Sec.~\ref{backtransf}. 
Then, it continues its propagation but, due to the translation it 
suffered while displacing 
the first imprint, it encodes its information into a displaced location. This 
complicated pulse dynamic is displayed in Fig.~\ref{pulse2}(b). 

Regardless of which situation took place, the location of the two imprints is 
now given by,
\begin{subequations}
\begin{align}
\kappa_a x_2^a & =\eta_{12}^a+\sigma\delta^{ab},\\
\kappa_b x_2^b & =\eta_{12}^b-\sigma\delta^{ab}.
\end{align}
\end{subequations}
An example of the resulting imprints after the two pulses have stored 
their information is presented in 
Fig.~\ref{den2}. The vertical dashed lines show the predicted location 
given by Eq.~\eqref{ximp}: These would be the actual locations of each 
imprint if they were done separately.

\subsection{Simultaneous imprint control\\}

\begin{figure}
\centering
\includegraphics[scale=1]{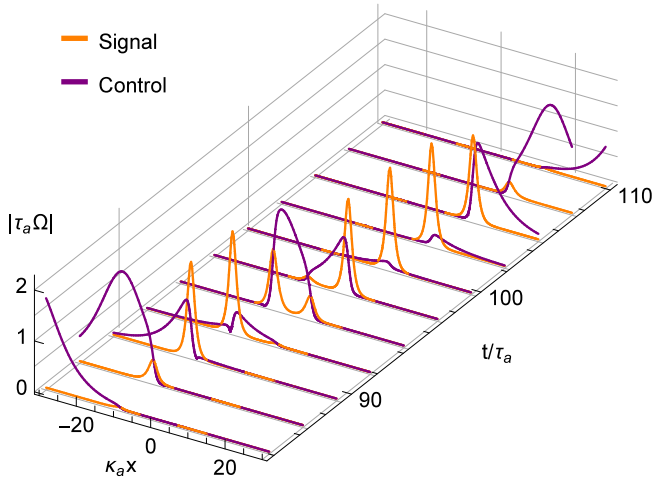}
\caption{\label{pulse3} Displacement step of the third order 
solution computed from the superposition of two type 1 and a type 2 first 
order solutions. The resulting imprints are depicted in Fig.~\ref{den3}.}
\end{figure}

Now that we have demonstrated that the encoding of two signal pulses is 
possible and have quantified the effect on each imprint due to the 
presence of the other, we need to address the question of 
multi-imprint control. To do this 
we consider the concrete example of pushing the imprints made by the two 
imprint solution discussed in the previous section via a control pulse 
of duration $\tau_c$. We compute this third order solution by means of the 
superposition rule [Eq.~\eqref{nls3}]. The resulting pulse dynamics for the 
displacement step are shown in Fig.~\ref{pulse3}. As the control pulse comes 
in, it first encounters the imprint made by the signal pulse $\tau_b$. 
Consequent to the interaction the signal pulse is retrieved, which 
in turn causes it to
propagate and start encoding its information back into the medium. The 
storage gives way to another control pulse which immediately starts 
interacting with the imprint left by the signal pulse $\tau_a$. This signal 
pulse is then retrieved, it propagates and finally is re-encoded into the 
medium at a displaced location giving way to a control pulse that propagates 
away at the speed of light.

\begin{figure}[t]
\centering
\includegraphics[scale=0.9]{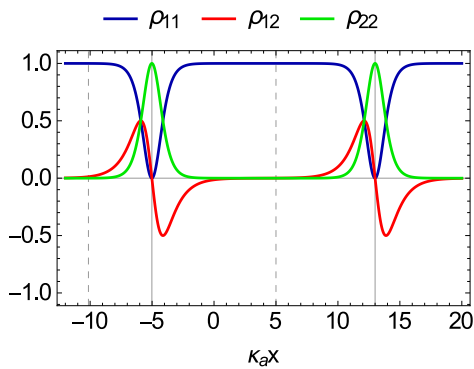}
\caption{\label{den3} Imprints displaced by a control pulse of 
duration such that $\tau_a>\tau_c>\tau_b$. The vertical continuous lines give 
the location of the imprints and the dashed ones show 
their previous location. 
These correspond to the pulse dynamics presented in Fig.~\ref{pulse3}.}
\end{figure}

Figure \ref{den3} shows the displaced imprint: The vertical dashed lines 
represent the location of each imprint before collision with the control 
pulse. There we can clearly see that each imprint was displaced a different 
amount and that just one of them suffered a $\pi$-phase shift. 
Each imprint is 
displaced according to their own parameters, that is, the new location for 
the $\tau_a$ imprint is
\begin{equation}
\kappa_ax_3^a=\kappa_a x_2^a +\delta^{ac}
\end{equation}
and for the $\tau_b$ imprint we have
\begin{equation}
\kappa_bx_3^b=\kappa_b x_2^b +\delta^{bc}.
\end{equation}
As we considered the case $\tau_a>\tau_c>\tau_b$ for plotting Figs. 7 and 8, 
following the 
results previously stated, we get that only the imprint with larger
 duration suffers the 
$\pi$-phase shift. Therefore we have shown that the manipulation of multiple 
imprints follows the single imprint rules as long as one respects the spatial 
limits of each imprint. That means that the control pulse should not push the 
first imprint it encounters beyond the second one. This does not imply that 
it cannot be done, just that the end result for the imprints is going to be 
different. From everything that has been said so far it is not hard to say 
what would happen in this scenario. When the control pulse collides with the 
first imprint it retrieves the signal pulse stored. This signal pulse then 
encounters the second imprint and thus interacts with it as a 
backward-transfer solution displacing the imprint to the left. 
Finally the signal pulse encodes its information, thus effectively 
inverting the order of the imprints. 

If instead of moving the imprints with a control pulse we had chosen a 
type 3 first order solution then everything would have been reversed. The 
imprints would be moved to the left according to the results derived in 
Sec.~\ref{backtransf}. Here again, there is an option for
 inverting the order 
of the imprints. If the duration of the type 3 pulse is tailored so that the 
imprint on the right is displaced more that the separation between imprints 
plus their widths, then the order is reversed. 
Each imprint would be moved in a 
different direction as well: The imprint on the right is displaced to the 
left by the signal pulse and the imprint on the left is pushed to the right 
by the control pulse that mediates the backward-transfer.\\

\section{Conclusions\\}

Throughout this manuscript we have shown the usefulness of a solution method 
via the Darboux transformation and the nonlinear superposition principle for 
generating sequences of pulses useful for storage and manipulation of 
information. It can be employed to compute higher order solutions which give 
rise to complicated sequences of pulses by purely algebraic, albeit tedious, 
calculations.

The new backward-transfer solution along with the solution presented in 
\citep{groves2013jaynes} allow complete control of the information encoded by 
a signal pulse in the ground state coherence of the atomic system. The 
imprint can be moved backward or forward any number of times by means of 
control and signal pulses. This is particularly important for unidirectional 
systems where the pulses can only propagate through the atomic medium in one 
direction, be that by design or experimental necessity. The generation of 
higher order solutions showed that the control of the imprint can be extended 
to any number of steps moving the imprint back and forth. In practice this 
might not be completely true: We will need to abandon the 
idealized condition of infinitely long pulses and media and 
consider the effect of decoherence due to 
spontaneous emission. But as existing numerical experiments show 
\cite{gutierrez2015manipulation}, storage and manipulation are still 
possible with a somewhat different dependence on the different parameters 
than the analytical solution, and they still present the 
same trend. Of course the degrading effects of decoherence 
may limit the number of steps in order 
to keep a certain degree of fidelity in the information stored. Note that 
the finiteness of the medium provides a way to retrieve the signal pulse by 
frustrating its re-encoding by the end face when displaced by a control 
pulse.

We also showed that multi-pulse storage is possible, and the effects due to 
the presence of another imprint can be quantified. This study could be 
continued to analyze the effect of encoding more pulses and see if the 
locations of the imprints follow a predictable trend from which it would be 
possible to extrapolate the behavior for any number of imprints. 
Additionally, we showed that manipulation of the imprints by means 
of pulses of type 2 and 3 is possible. In this case, there are 
additional consideration, such 
as the spatial extent of each imprint. Overlapping the imprints must be 
avoided when displacing the imprint a long distance or inverting their 
order. But from the numerical data, there is a maximum displacement that 
could very well hinder any chance of inverting the imprint or overlapping 
them (as long as enough room was left between them during the encoding stage) 
for any realistic experimental scenario.

A final note must be made about the effects of Doppler broadening on the encoding and retrieval of the signal pulse. In Ref.~\cite{clader2007two} Clader and Eberly worked out the first order solution presented here with the added effects of Doppler broadening. From their work we can see that, in the case where the Doppler distribution is centered around resonance, the imprint made on all the atoms is located at the same place but (assuming a real coherence $\rho_{12}$ on resonance) for the atoms off resonance the real part of the ground-state coherence is attenuated and acquires an imaginary part. This imaginary part has a sign that depends on the sign of the detuning, so there will be an equal number of atoms with negative and positive imaginary parts. When taking the average over the Doppler distribution, this contribution will cancel out. Therefore, when a control pulse collides with the imprint, the only consequence is the attenuation of the real part for some atoms which will inevitably hinder the retrieval of the signal pulse but never suppress it completely.   \\

\section*{Acknowledgements\\}
This research was funded by the National Science Foundation (NSF) (PHY-1203931, PHY-1505189) and
 RGC acknowledges the support of a CONACYT fellowship.

\end{document}